\documentclass[12pt,preprint]{aastex}
\begin{document}
\newcommand{\calu}{{\cal U}}
\newcommand{\calq}{{\cal Q}}
\newcommand{\bx}{{\rm \bf x}}
\newcommand{\bk}{{\bar{\kappa}}}
\title{Strong Lensing Probabilities in a Cosmological Model with
 a Running Primordial Power Spectrum}
\author{Tong-Jie Zhang}
\affil{Department of Astronomy, Beijing Normal University, Beijing
100875, P.R.China; tjzhang@bnu.edu.cn}
\author{Zhi-Liang Yang}
\affil{Department of Astronomy, Beijing Normal University, Beijing
100875, P.R.China; zlyang@bnu.edu.cn}
\author{Xiang-Tao He}
\affil{Department of Astronomy, Beijing Normal University, Beijing
100875, P.R.China; xthe@bnu.edu.cn}

\begin{abstract}
The combination of the first-year Wilkinson Microwave Anisotropy
Probe (WMAP) data with other finer scale cosmic microwave
background (CMB) experiments (CBI and ACBAR) and two structure
formation measurements (2dFGRS and Lyman $\alpha$ forest) suggest
a $\Lambda$CDM cosmological model with a running spectral power
index of primordial density fluctuations. Motivated by this new
result on the index of primordial power spectrum, we present the
first study on the predicted lensing probabilities of image
separation in a spatially flat $\Lambda$CDM model with a running
spectral index (RSI-$\Lambda$CDM model). It is shown that the
RSI-$\Lambda$CDM model suppress the predicted lensing
probabilities on small splitting angles of less than about
4$^{''}$ compared with that of standard power-law $\Lambda$CDM
(PL-$\Lambda$CDM) model.

\end{abstract}

\keywords{cosmology:theory---dark
matter---galaxies:halos---gravitational lensing---early
universe---large-scale structure\\
PACS numbers:98.80.Es,98.62.Sb,95.35.+d,98.80.Cq}

\section{Introduction}

Mapping the mass distribution of matter in the universe has been a
major challenge for modern observational cosmology. One of the
direct procedures to weigh matter in the universe is measuring its
deflection of light by gravity. The statistics of gravitational
lensing can provide us with a very powerful probe of the mass
distribution of the Universe. It is well known that the
Jodrell-Bank VLA Astrometric Survey (JVAS) and the Cosmic Lens
All-Sky Survey (CLASS)
\citep{2000IAUS..201E..47B,2003MNRAS.341....1M,
2003MNRAS.341...13B} have provided us with observations of strong
lensing probabilities for small image separations ranging from
$0.3''$ to $3''$. By comparing predicted lensing probabilities
with observations, we can examine the mass distributions of dark
matter halos, in particular, their inner density slopes
\citep{2004ApJ...600L...7H,2003MNRAS.344.1237R}.

Based on the Cold Dark Matter (CDM) model, which has become the
standard theory of cosmic structure formation, the lensing
probabilities strongly depend on the density profiles of CDM
halos. The lensing model is usually described by a singular
isothermal sphere (SIS), the Navarro-Frenk-White (NFW) model
\citep{1996ApJ...462..563N,1997ApJ...490..493N}, or generalized
NFW (GNFW) density profiles of dark halos
\citep{1996MNRAS.278..488Z}. \cite{2002ApJ...566..652L} employed a
semi-analytical approach to analyze the gravitational lensing of
remote quasars by foreground dark halos and checked the
plausibility of various lensing models. They found that no model
can completely explain the current observations: the SIS models
predict too many lenses with large splitting angles, while the NFW
models predict too few small splitting angles. They therefore
further developed a two-population halo model for lensing: small
mass halos with a steep inner density slope and large mass halos
with a shallow inner density slope, concluding that a combination
of SIS and NFW halo models can reproduce the current observations
reasonably well. Unlike previous work that directly models the
density profiles of dark matter halos semi-analytically,
\cite{2004ApJ...602L...5Z} generalized the density profiles of
dark matter halos from high-resolution N-body simulations by means
of generalized Navarro-Frenk-White (GNFW) models of three
populations with slopes, $\alpha$, of about -1.5, -1.3 and -1.1
for galaxies, groups and clusters, respectively. He presented the
calculations of lensing probabilities using these GNFW profiles
for three populations in various spatially flat cosmological
models with a cosmological constant $\Lambda$. He showed that the
compound model of density profiles does not match well with the
lensing probabilities derived from the combined data of
JVAS/CLASS. Recently, \cite{2004astro.ph..7298M} compared
predictions on small scale structure for the $\Lambda$CDM model by
numerical simulations with observed flux ratios, and found that
the disagreements between monochromatic flux ratios and simple
lens models can be explained without any substructure in the dark
matter halos of primary lenses. However, spectroscopic lensing
observations of Q2237+0305 require more small mass dark halos than
that expected in the $\Lambda$CDM model.

In particular, a power spectrum of primordial fluctuation,
$P_p(k)$, should be assumed in advance in the calculation of
lensing probabilities. Inflationary models predict a approximately
scale-invariant power spectra for primordial density (scalar
metric) fluctuation, $P_p(k)\propto k^n$ with index $n=1$
\citep{1982PhRvL..49.1110G,1983PhRvD..28..679B}. The combination
of the first-year Wilkinson Microwave Anisotropy Probe (WMAP) data
with other finer scale cosmic background (CMB) experiments (Cosmic
Background Imager [CBI], Arcminute Cosmology Bolometer Array
Receiver [ACBAR]) and two observations of large-scale structure
(the Anglo-Australian Telescope Two-Degree Field Galaxy Redshift
Survey [2dFGRS] and Lyman $\alpha$ forest) favor a $\Lambda$CDM
cosmological model with a running index of the primordial power
spectrum (RSI-$\Lambda$CDM), while the WMAP data alone still
suggest a best-fit standard power-law $\Lambda$CDM model with the
spectral index of $n\approx 1$ (PL-$\Lambda$CDM)
\citep{2003ApJS..148..175S,2003ApJS..148..213P}. However, there
still exist the intriguing discrepancies between theoretical
predictions and observations on both the largest and smallest
scales. While the emergence of a running spectral index may
improve problems on small scales, there remain a possible
discrepancy on the largest angular scales. It is particularly
noted that the running spectral index model suppress significantly
the power amplitude of fluctuations on small scales
\citep{2003ApJS..148..175S,2003ApJ...598...73Y}. This imply a
reduction of the amount of substructure within galactic halos
\citep{2002PhRvD..66d3003Z}. \cite{2003ApJ...598...73Y} studied
early structure formation in a RSI-$\Lambda$CDM universe using
high-resolution cosmological N-body/hydrodynamic simulations. They
showed that the reduced small-scale power in the RSI-$\Lambda$CDM
model causes a considerable delay in the formation epoch of
low-mass minihalos ($\sim10^6 M_{\sun}$) compared with the
PL-$\Lambda$CDM model, although early structure still forms
hierarchically in the RSI-$\Lambda$CDM model. Thus the running
index probably affect the strong lensing process of distant
sources by intervening dark matter halos because the lensing
probabilities strongly depend on the abundance of dark halos
formed in the evolution of the universe. In this letter, we will
present the first calculation of lensing probabilities in a
RSI-$\Lambda$CDM model to explore the effect of running spectral
index of primordial fluctuation on strong lensing probabilities.
We adopt a spatially flat $\Lambda$CDM cosmological model through
this letter.

The reminder of this paper is organized as follows. We describe
strong lensing probabilities and mass function of dark halos in
Section 2 and 3 respectively. The cross-sections for producing
multiple images are described in Section 4. Our results is shown
in Section 5, while a conclusion and discussion are given in
Section 6.

\section{Strong Lensing Probability}

The lensing probability with image separation larger than
$\Delta\theta$ is given by \cite{1992grle.book.....S}
\begin{equation}
P(>\Delta\theta)=\int p(z_s)dz_s
\int^{z_{\mathrm{s}}}_0\frac{dD_{\mathrm{p}}(z)}
{dz}dz\int^{\infty}_0\bar{n}(M,z)\sigma(M,z)dM. \label{prob}
\end{equation}
In this expression, $D_{\mathrm{p}}(z)=c/H_0\int^z_0 dz/(1+z)E(z)$
is the proper distance from the observer to the lens at redshift
$z$ where the expansion rate of the universe
$E(z)=\sqrt{\Omega_\mathrm{m} (1+z)^3+\Omega_{\Lambda}}$ in a
spatially flat cosmological model, and $p(z_s)$ is the redshift
distribution of distant sources. The physical number density
$\bar{n}(M,z)$ of virialized dark halos of masses between $M$ and
$M+dM$ is expressed as $\bar{n}(M,z)=n(M,z)(1+z)^3$ and
$\sigma(M,z)$ is the cross-section defined in the lens plane for
forming multiple images.

\section{Mass Function of Dark Halos}

In the standard hierarchical theory of structure formation, the
comoving number density of virialized dark halos per unit mass $M$
at redshift $z$ can be given by the Press and Schechter (PS)
formula \citep{1974ApJ...187..425P}: $n(M,z)=dN/dM=\rho_0
f(M,z)/M$ where $\rho_0$ is the mean mass density of the universe
today and, instead of PS formula in this letter, the mass function
$f(M,z)$ takes the form of an empirical fit from high-resolution
simulation \citep{2001MNRAS.321..372J}
\begin{equation}
f(M,z)=\frac{0.301}{M}\frac{d\ln\Delta^{-1}(M,z)}{d\ln
M}\exp(-|\ln\Delta^{-1}(M,z)+0.64|^{3.88}).
\label{jenkins}
\end{equation}
Here, $\Delta(M,z)=\Delta(M)D(z)$ and
$D(z)=e(\Omega(z))/e(\Omega_{\mathrm{m}})(1+z)$ is the linear
growth function of density perturbation
\citep{1992ARA&A..30..499C}, in which
$e(x)=2.5x/(1/70+209x/140-x^2/140+x^{4/7})$ and
$\Omega(z)=\Omega_{\mathrm{m}}(1+z)^3/ E^2(z)$. The present
variance of the fluctuations within a sphere containing a mass $M$
can be expressed as
$\Delta^2(M)=\frac{1}{2\pi^2}\int^{\infty}_0P(k)
W^2(kr_{\mathrm{M}})k^2dk$, where
$W(kr_{\mathrm{M}})=3[\sin(kr_{\mathrm{M}})/(kr_{\mathrm{M}})^3-
\cos(kr_{\mathrm{M}})/(kr_{\mathrm{M}})^2]$ is the Top-hat window
function in Fourier space and $
r_{\mathrm{M}}=(3M/4\pi\rho_0)^{1/3}$. The power spectrum of CDM
density fluctuations is $P(k)=P_p(k)T^2(k)$ where the matter
transfer function $T(k)$ is given by \cite{1999ApJ...511....5E},
and $P_p(k)$ is the primordial power spectrum of density
fluctuation. The scale-invariant primordial power spectrum in the
PL-$\Lambda$CDM model is given by $P_p(k)=Ak^{n_s}$ with index
$n_s$=1 and that in the RSI-$\Lambda$CDM model is assumed to be
$P_p(k)=P(k_0)(k/k_0)^{n_{s}(k)}$, where the index $n_{s}(k)$ is a
function of length scale
\begin{equation}
n_{s}(k)=n_{\rm s}(k_0) +\frac{1}{2}\frac{{\rm d}n_{s}(k)}{{\rm
d}\ln k}\ln \left(\frac{k}{k_0}\right).
\end{equation}
The pivot scale $k_0$=0.05 h Mpc$^{-1}$, $n_{\rm s}(k_0)$=0.93,
and $dn_{s}/d\ln k$=-0.03 are the best-fit values to the
combination data of the recent CMB experiments and two other
large-scale structure observations \citep{2003ApJS..148..175S}.
For both PL-$\Lambda$CDM and RSI-$\Lambda$CDM models, the
amplitude of primordial power spectrum, $A$ and $P(k_0)$, are
normalized to $\sigma_8=\Delta
(r_{\mathrm{M}}=8h^{-1}\mathrm{Mpc})$, which is the rms mass
fluctuations when present universe is smoothed using a window
function on a scale of $8h^{-1}\mathrm{Mpc}$.

\section{The Cross-sections for Producing Multiple Images}

The cross-section for producing multiple images relies on the
density profile of dark matter halos. Based on previous work
\citep{2002ApJ...566..652L,2004ApJ...602L...5Z}, we model dark
halos as an SIS for $M<M_{c1}$ and NFW profile for $M>M_{c1}$,
respectively, where $M_{c1}\sim10^{13}h^{-1}M_{\sun}$ corresponds
to the cooling mass scale
\citep{2000ApJ...532..679P,2001ApJ...559..531K}.

\subsection{The Cross-section for SISs}

The mass density for a lens of SIS is $\rho(r)=\sigma_v^2/2\pi G
r^2$ where $\sigma_v$ is the velocity dispersion
\citep{1992grle.book.....S}. Thus the surface mass density of the
SIS satisfies $\Sigma(\xi)= \sigma_v^2/2G\xi$ where
$\xi\equiv|\vec{\xi}|$, $\vec{\xi}$ is the position vector in the
lens plane. By defining the length scales in the lens plane and
the source plane as $\xi_0=4\pi(\sigma_v/c)^2d^A_L
d^A_{LS}/d^A_S\,,\eta_0=\xi_{0}d^A_S/d^A_L$ respectively, we can
simplify the lensing equation and obtain the image separation of
SIS lensing
\begin{equation}
\Delta\theta=1.27^{\prime}({d^A_{LS}\over
d^A_S})M_{15}^{2/3}E(z)^{2/3},
\end{equation}
where $M_{15}=M/(10^{15}\mathrm{h} ^{-1}M_{\sun})$ is
dimensionless halo mass. Here $d^A_S$ and $d^A_L$ are the angular
diameter distances from the observer to the source and to the lens
object respectively, while $d^A_{LS}$ is the same quantity but
from the lens to the source object. Therefore, the cross-section
defined in the lens plane for forming two images by an SIS lens
with splitting angle $\Delta\theta
>\Delta\theta_0$ can be expressed as $\sigma(M,z)=\pi\xi_0^2\vartheta
(\Delta\theta-\Delta\theta_0)$ where $\vartheta$ is a step
function.

\subsection{The Cross-section for NFW Models}

The GNFW density profile can be expressed in the form
$\rho(r)={\rho_s r_s^3/r^{\alpha}(r+r_s)^{3-\alpha}}$
\citep{1996MNRAS.278..488Z} where $0<\alpha<3$. The GNFW density
profile reduces to the case of NFW if $\alpha=1$.  The mass of a
dark halo within $r_{200}$ can be defined as
$M=4\pi\int^{r_{200}}_0\rho
r^2dr=4\pi\rho_\mathrm{s}r_\mathrm{s}^3f(c_1)$, and $r_{200}$ is
the radius of a sphere around a dark halo within which the average
mass density is $200$ times the critical mass density of the
universe. The function $f(c_1)=\int_0^{c_1} {x^2 dx/x^\alpha
(1+x)^{3-\alpha}}$ and $c_1=r_{200}/r_\mathrm{s}$ is the
concentration parameter and takes the form
\citep{2001ApJ...559..572O,2002ApJ...568..488O,2004astro.ph..8573O}
\begin{eqnarray}
    c_1(M_{15},z)=c_{\rm norm}
    \frac{2-\alpha}{1+z}[10M_{15}]^{-0.13},
    \label{c1}
\end{eqnarray}
where $z$ is the redshift of halo, $c_{\rm norm}=8$
\citep{2001MNRAS.321..559B}. So $\rho_\mathrm{s}$ and
$r_\mathrm{s}$ can be related to mass $M_{15}$ and redshift $z$ by

\begin{equation}
\rho_\mathrm{s}=\rho_\mathrm{c_0}
E^2(z)\frac{200}{3}\frac{c_1^3}{f(c_1(M_{15},z))},\
r_\mathrm{s}=\frac{1.626}{c_1}\frac{M_{15}^{1/3}}
{E^{2/3}(z)}h^{-1}\mathrm{Mpc},
\end{equation}
where $\rho_{c_0}$ is the critical mass density of the universe
today. The lensing equation for the GNFW profile is given by
$y=x-\mu_s g(x)/x$; $\vec{\xi}=\vec{x}r_s$ and
$\vec{\eta}=\vec{y}r_s d^A_S/d^A_L$ are the position vectors in
the lens plane and the source plane respectively, $g(x) \equiv
\int_0^x u du \int_0^{\infty} \left(u^2 +z^2\right)^{-\alpha/2}
\left[\left(u^2+z^2\right)^{1/2} +1\right]^{-3+\alpha} dz$, and
\begin{equation}
\mu_s\equiv{4\rho_s r_s\over \Sigma_{\rm
cr}}=0.002(\frac{\rho_s}{\rho_{c_0}})({r_s\over 1 h^{-1} {\rm
Mpc}})({d^A_R\over c/H_0}), \label{alps}
\end{equation}
where $\mu_s$ is a parameter on which the efficiency of producing
multiple images is strongly dependent. Here $\Sigma_{\rm
cr}={c^2\over 4\pi G}\,{d^A_S\over d^A_L d^A_{LS}}$ is the
critical surface mass density, and $d^A_R=d^A_L d^A_{LS}/d^A_S$.
The lensing equation curves are symmetrical with respect to the
origin. Multiple images can be formed when $|y|\leq y_{\rm cr}$,
where $y_{\rm cr}$ is the maximum value of $y$ when $x<0$ or the
minimum value for $x>0$. Generally speaking, there exist three
images for $|y|< y_{\rm cr}$. We will just consider the outermost
two images stretched by the splitting angle $\Delta\theta$ when
more than two images are formed. Therefore, we can write the
cross-section as $\sigma\left(M,z\right) \approx \pi y_{\rm cr}^2
r_s^2\,\vartheta\left(\Delta\theta - \Delta\theta_0\right)$ with
$\Delta\theta>\Delta\theta_0$ in the lens plane for multiple
images produced by a GNFW lens at $z$. Here splitting angle
$\Delta\theta$ is given by $\Delta\theta={r_s\Delta x/d^A_L}
\approx{2 x_0 r_s/d^A_L}$ and $x_0$ is the positive root of the
lensing equation $y(x)=0$.

\section{Numerical Results}
In this letter, we assume spatially flat $\Lambda$CDM models
characterized by the matter density parameter $\Omega_{\mathrm
m}$, vacuum energy density parameter $\Omega_{\Lambda}$. For both
PL-$\Lambda$CDM and RSI-$\Lambda$CDM models, we take cosmological
parameters to be the new result from the WMAP: Hubble constant
$h=0.71$, $\Omega_{\mathrm m}=0.27$, $\sigma_8=0.84$
\citep{2003ApJS..148....1B,2003ApJS..148..175S} and
$M_{c1}=1.5\times10^{13}h^{-1}M_{\sun}$. As mentioned above, the
lensing probability depend strongly on the abundance of viralized
dark halos. The mass function of dark halos directly involve the
calculation of primordial power of density fluctuation. According
to Eq.(\ref{prob}), it is clear that the effect of running
spectral index on mass function of halos cause the difference of
lensing probabilities between the two models.

\begin{figure}
\epsscale{0.8} \plotone{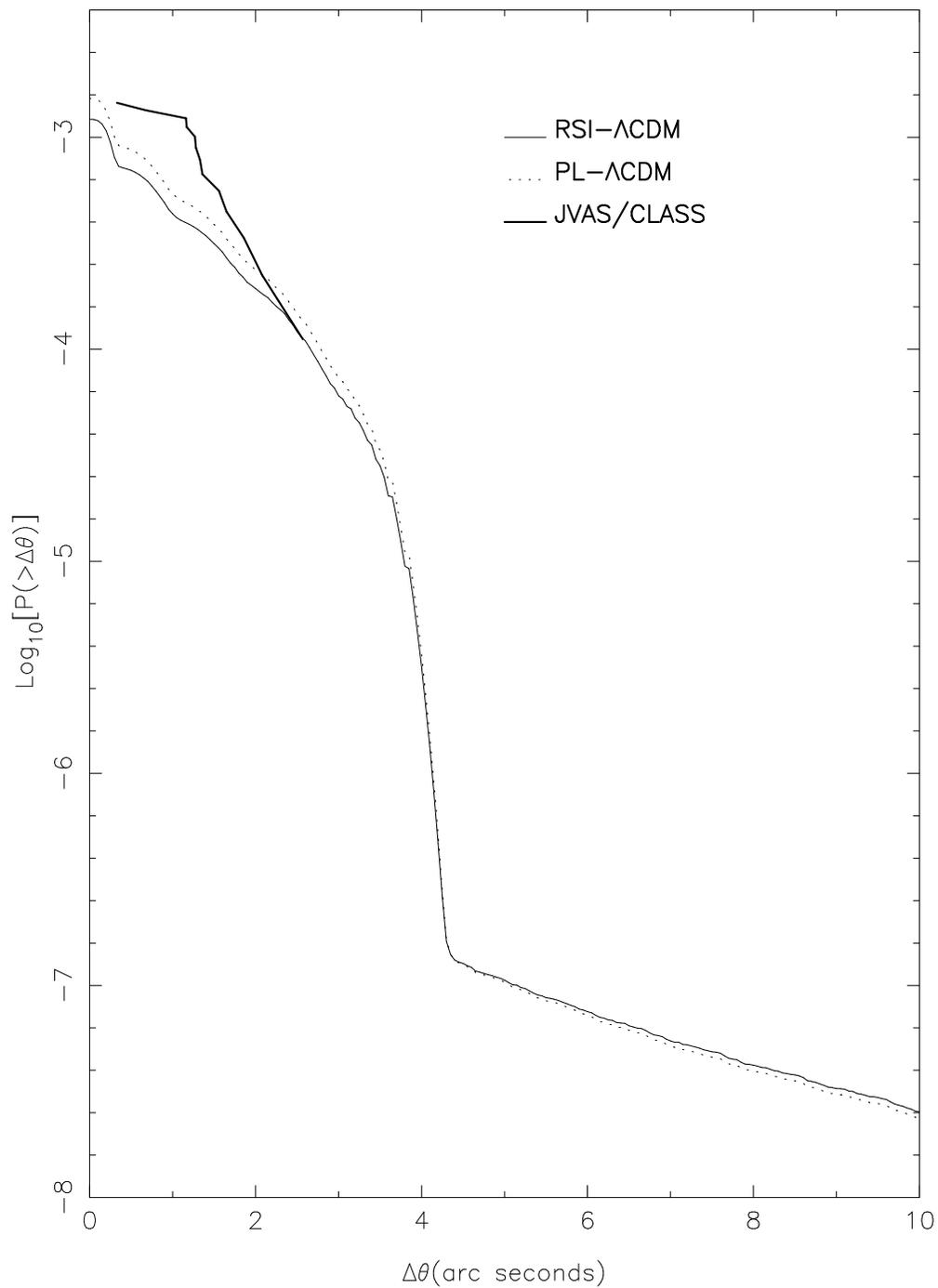} \caption{The lensing probabilities
for image separations greater than $\Delta\theta$. The solid-line
histogram represents observed lensing probabilities from
JVAS/CLASS. The solid line is the lensing probabilities for
running spectral index $\Lambda$CDM model, while the dashed line
is that for power law $\Lambda$CDM model.} \label{f2eps}
\end{figure}

Although the redshift distribution of quasars in the JVAS/CLASS
survey is still poorly known, the prediction of
\cite{1990MNRAS.247...19D} model and the CLASS lensing sub-sample
redshift measurements suggest that the redshift distribution for
CLASS unlensed sources can be modelled by a Gaussian distribution
with mean redshift $<z_s>$=1.27 \citep{2000AJ....119.2629M} and
dispersion $\sigma_z$=0.95
\citep{2002PhRvL..89o1301C,2003MNRAS.341...13B,2003MNRAS.341....1M}.
Thus in this letter we adopt the Gaussian redshift distribution of
quasars with the mean redshift $<z_s>$=1.27 and dispersion
$\sigma_z$=0.95. In the definition of cross-section for forming
two images, we just consider the criterion $\Delta\theta$, and
neglect another one $q_r$ that is the flux density (brightness)
ratio of multiple images \citep{1992grle.book.....S}. In order to
investigate the effect of central black holes or bulges on lensing
probability, \cite{2003ApJ...587L..55C,2003A&A...397..415C}
introduced $q_r$ into the calculation for lensing cross-section.
Due to the existence of central black holes or galactic bulges,
$y_{cr}$ becomes extremely large when $|x|$ approaches zero. Thus
$y_{cr}$ can be determined by the consideration of $q_r$ together
with $\Delta\theta$. However for GNFW halo models in the absence
of central black holes or galactic bulges, the lensing equation
curves are so smooth that we do not need to define cross-section
by $q_r$. As for SISs, it is necessary to introduce $q_r$ because
both the predicted lensing probabilities and the determination of
the cosmic equation of state $\omega$ are quite sensitive to $q_r$
\citep{2004ChJAA...4..118C,2004A&A...418..387C}. Our objective in
this letter is to
examine the effect of running spectral index on the predicted
lensing probabilities, so for explicit we neglect this selection
criterion $q_r$ for forming multiple images. In such cases, we
calculate lensing probabilities with image separations greater
than $\Delta\theta$ according to the compound model of SIS and NFW
halo profile in both PL-$\Lambda$CDM and RSI-$\Lambda$CDM models.
Our numerical results are shown in Fig.\ref{f2eps} together with
the observational one from JVAS/CLASS. As we expect, there is
slight difference between the two models on large image
separations, while this difference enlarges with the decrease of
the splitting angle. More specifically, the RSI-$\Lambda$CDM model
can reduce the predicted lensing probabilities on small splitting
angle of less than about 4$^{''}$ compared with that of
PL-$\Lambda$CDM model.

\section{Conclusions and Discussion}

Motivated by the new result on the index of primordial power
spectrum from a combination of WMAP data with other finer scale
CMB experiments and other large-scale structure observations, we
present the first study on the predicted lensing probability in a
$\Lambda$CDM model with a running spectral index. In a popular
cosmological model from the new fit results mentioned above:
$h=0.71$, $\Omega_{\mathrm m}=0.27$, $\sigma_8=0.84$, we calculate
lensing probabilities with image separations greater than
$\Delta\theta$ according to the compound model of SIS and NFW halo
profile in both PL-$\Lambda$CDM and RSI-$\Lambda$CDM models. From
the analysis above, we can see that the running spectral index
mainly affect the predicted lensing probability on small image
separation. It is well known that structures in the universe forms
hierarchically in standard CDM models. \cite{2003ApJ...598...73Y}
found that although this hierarchical formation mechanism do not
work well in RSI-$\Lambda$CDM model compared with that in
PL-$\Lambda$CDM model and it also is not clear that the PS theory
can be used in RSI-$\Lambda$CDM model, the mass function measured
by high-resolution cosmological N-body/hydrodynamic simulations
overall match the PS mass function for both RSI-$\Lambda$CDM and
PL-$\Lambda$CDM model.

In addition, because the running spectral index model predicts a
significant lower power of density fluctuation on small scales
than the standard PL-$\Lambda$CDM model
\citep{2003ApJS..148..175S,2003ApJ...598...73Y}, it should also
attract considerable attention in studies on weak lensing by
large-scale structure \citep{2003astro.ph..8446I}, especially on
skewness  \citep{2003ApJ...592..664P,2003ApJ...598..818Z}
 which characterizes the non-Gaussian property of $\kappa$
field in the nonlinear regime.

We are very grateful to the anonymous
referee for constructive suggestion. T.J.Zhang would like to thank
Ue-Li Pen, Peng-Jie Zhang, Xiang-Ping Wu, Bo Qin and CITA for
their hospitality during his visits to
the Canadian
Institute for Theoretical Astrophysics(CITA), University of
Toronto and the cosmology groups of the
National Astronomical Observatories of P.R.China.
This work was supported by the National Science Foundation of
China (grants No.10473002 and 10273003).
\bibliography{ztjcos-lensbib}

\begin{thebibliography}{39}
\expandafter\ifx\csname natexlab\endcsname\relax\def\natexlab#1{#1}\fi

\bibitem[{{Bardeen} {et~al.}(1983){Bardeen}, {Steinhardt}, \&
  {Turner}}]{1983PhRvD..28..679B}
{Bardeen}, J.~M., {Steinhardt}, P.~J., \& {Turner}, M.~S. 1983, \prd, 28, 679

\bibitem[{{Bennett} {et~al.}(2003){Bennett}, {Halpern}, {Hinshaw}, {Jarosik},
  {Kogut}, {Limon}, {Meyer}, {Page}, {Spergel}, {Tucker}, {Wollack}, {Wright},
  {Barnes}, {Greason}, {Hill}, {Komatsu}, {Nolta}, {Odegard}, {Peiris},
  {Verde}, \& {Weiland}}]{2003ApJS..148....1B}
{Bennett}, C.~L., {Halpern}, M., {Hinshaw}, G., {Jarosik}, N., {Kogut}, A.,
  {Limon}, M., {Meyer}, S.~S., {Page}, L., {Spergel}, D.~N., {Tucker}, G.~S.,
  {Wollack}, E., {Wright}, E.~L., {Barnes}, C., {Greason}, M.~R., {Hill},
  R.~S., {Komatsu}, E., {Nolta}, M.~R., {Odegard}, N., {Peiris}, H.~V.,
  {Verde}, L., \& {Weiland}, J.~L. 2003, \apjs, 148, 1

\bibitem[{{Browne} \& {Myers}(2000)}]{2000IAUS..201E..47B}
{Browne}, I.~W.~A. \& {Myers}, S.~T. 2000, in IAU Symposium

\bibitem[{{Browne} {et~al.}(2003){Browne}, {Wilkinson}, {Jackson}, {Myers},
  {Fassnacht}, {Koopmans}, {Marlow}, {Norbury}, {Rusin}, {Sykes}, {Biggs},
  {Blandford}, {de Bruyn}, {Chae}, {Helbig}, {King}, {McKean}, {Pearson},
  {Phillips}, {Readhead}, {Xanthopoulos}, \& {York}}]{2003MNRAS.341...13B}
{Browne}, I.~W.~A., {Wilkinson}, P.~N., {Jackson}, N.~J.~F., {Myers}, S.~T.,
  {Fassnacht}, C.~D., {Koopmans}, L.~V.~E., {Marlow}, D.~R., {Norbury}, M.,
  {Rusin}, D., {Sykes}, C.~M., {Biggs}, A.~D., {Blandford}, R.~D., {de Bruyn},
  A.~G., {Chae}, K.-H., {Helbig}, P., {King}, L.~J., {McKean}, J.~P.,
  {Pearson}, T.~J., {Phillips}, P.~M., {Readhead}, A.~C.~S., {Xanthopoulos},
  E., \& {York}, T. 2003, \mnras, 341, 13

\bibitem[{{Bullock} {et~al.}(2001){Bullock}, {Kolatt}, {Sigad}, {Somerville},
  {Kravtsov}, {Klypin}, {Primack}, \& {Dekel}}]{2001MNRAS.321..559B}
{Bullock}, J.~S., {Kolatt}, T.~S., {Sigad}, Y., {Somerville}, R.~S.,
  {Kravtsov}, A.~V., {Klypin}, A.~A., {Primack}, J.~R., \& {Dekel}, A. 2001,
  \mnras, 321, 559

\bibitem[{{Carroll} {et~al.}(1992){Carroll}, {Press}, \&
  {Turner}}]{1992ARA&A..30..499C}
{Carroll}, S.~M., {Press}, W.~H., \& {Turner}, E.~L. 1992, \araa, 30, 499

\bibitem[{{Chae} {et~al.}(2002){Chae}, {Biggs}, {Blandford}, {Browne}, {de
  Bruyn}, {Fassnacht}, {Helbig}, {Jackson}, {King}, {Koopmans}, {Mao},
  {Marlow}, {McKean}, {Myers}, {Norbury}, {Pearson}, {Phillips}, {Readhead},
  {Rusin}, {Sykes}, {Wilkinson}, {Xanthopoulos}, \&
  {York}}]{2002PhRvL..89o1301C}
{Chae}, K.-H., {Biggs}, A.~D., {Blandford}, R.~D., {Browne}, I.~W., {de Bruyn},
  A.~G., {Fassnacht}, C.~D., {Helbig}, P., {Jackson}, N.~J., {King}, L.~J.,
  {Koopmans}, L.~V., {Mao}, S., {Marlow}, D.~R., {McKean}, J.~P., {Myers},
  S.~T., {Norbury}, M., {Pearson}, T.~J., {Phillips}, P.~M., {Readhead}, A.~C.,
  {Rusin}, D., {Sykes}, C.~M., {Wilkinson}, P.~N., {Xanthopoulos}, E., \&
  {York}, T. 2002, Physical Review Letters, 89, 151301

\bibitem[{{Chen}(2003{\natexlab{a}})}]{2003ApJ...587L..55C}
{Chen}, D. 2003{\natexlab{a}}, \apjl, 587, L55

\bibitem[{{Chen}(2003{\natexlab{b}})}]{2003A&A...397..415C}
---. 2003{\natexlab{b}}, \aap, 397, 415

\bibitem[{{Chen}(2004{\natexlab{a}})}]{2004ChJAA...4..118C}
---. 2004{\natexlab{a}}, Chinese Journal of Astronony and Astrophysics, 4, 118

\bibitem[{{Chen}(2004{\natexlab{b}})}]{2004A&A...418..387C}
{Chen}, D.-M. 2004{\natexlab{b}}, \aap, 418, 387

\bibitem[{{Dunlop} \& {Peacock}(1990)}]{1990MNRAS.247...19D}
{Dunlop}, J.~S. \& {Peacock}, J.~A. 1990, \mnras, 247, 19

\bibitem[{{Eisenstein} \& {Hu}(1999)}]{1999ApJ...511....5E}
{Eisenstein}, D.~J. \& {Hu}, W. 1999, \apj, 511, 5

\bibitem[{{Guth} \& {Pi}(1982)}]{1982PhRvL..49.1110G}
{Guth}, A.~H. \& {Pi}, S.-Y. 1982, Physical Review Letters, 49, 1110

\bibitem[{{Huterer} \& {Ma}(2004)}]{2004ApJ...600L...7H}
{Huterer}, D. \& {Ma}, C. 2004, \apjl, 600, L7

\bibitem[{{Ishak} {et~al.}(2003){Ishak}, {Hirata}, {McDonald}, \&
  {Seljak}}]{2003astro.ph..8446I}
{Ishak}, M., {Hirata}, C.~M., {McDonald}, P., \& {Seljak}, U. 2003, ArXiv
  Astrophysics e-prints:astro-ph/0308446

\bibitem[{{Jenkins} {et~al.}(2001){Jenkins}, {Frenk}, {White}, {Colberg},
  {Cole}, {Evrard}, {Couchman}, \& {Yoshida}}]{2001MNRAS.321..372J}
{Jenkins}, A., {Frenk}, C.~S., {White}, S.~D.~M., {Colberg}, J.~M., {Cole}, S.,
  {Evrard}, A.~E., {Couchman}, H.~M.~P., \& {Yoshida}, N. 2001, \mnras, 321,
  372

\bibitem[{{Kochanek} \& {White}(2001)}]{2001ApJ...559..531K}
{Kochanek}, C.~S. \& {White}, M. 2001, \apj, 559, 531

\bibitem[{{Li} \& {Ostriker}(2002)}]{2002ApJ...566..652L}
{Li}, L. \& {Ostriker}, J.~P. 2002, \apj, 566, 652

\bibitem[{{Marlow} {et~al.}(2000){Marlow}, {Rusin}, {Jackson}, {Wilkinson},
  {Browne}, \& {Koopmans}}]{2000AJ....119.2629M}
{Marlow}, D.~R., {Rusin}, D., {Jackson}, N., {Wilkinson}, P.~N., {Browne},
  I.~W.~A., \& {Koopmans}, L. 2000, \aj, 119, 2629

\bibitem[{{Metcalf}(2004)}]{2004astro.ph..7298M}
{Metcalf}, R.~B. 2004, ArXiv Astrophysics e-prints:astro-ph/0407298

\bibitem[{{Myers} {et~al.}(2003){Myers}, {Jackson}, {Browne}, {de Bruyn},
  {Pearson}, {Readhead}, {Wilkinson}, {Biggs}, {Blandford}, {Fassnacht},
  {Koopmans}, {Marlow}, {McKean}, {Norbury}, {Phillips}, {Rusin}, {Shepherd},
  \& {Sykes}}]{2003MNRAS.341....1M}
{Myers}, S.~T., {Jackson}, N.~J., {Browne}, I.~W.~A., {de Bruyn}, A.~G.,
  {Pearson}, T.~J., {Readhead}, A.~C.~S., {Wilkinson}, P.~N., {Biggs}, A.~D.,
  {Blandford}, R.~D., {Fassnacht}, C.~D., {Koopmans}, L.~V.~E., {Marlow},
  D.~R., {McKean}, J.~P., {Norbury}, M.~A., {Phillips}, P.~M., {Rusin}, D.,
  {Shepherd}, M.~C., \& {Sykes}, C.~M. 2003, \mnras, 341, 1

\bibitem[{{Navarro} {et~al.}(1996){Navarro}, {Frenk}, \&
  {White}}]{1996ApJ...462..563N}
{Navarro}, J.~F., {Frenk}, C.~S., \& {White}, S.~D.~M. 1996, \apj, 462, 563

\bibitem[{{Navarro} {et~al.}(1997){Navarro}, {Frenk}, \&
  {White}}]{1997ApJ...490..493N}
---. 1997, \apj, 490, 493

\bibitem[{{Oguri}(2004)}]{2004astro.ph..8573O}
{Oguri}, M. 2004, ArXiv Astrophysics e-prints:astro-ph/0408573

\bibitem[{{Oguri} {et~al.}(2001){Oguri}, {Taruya}, \&
  {Suto}}]{2001ApJ...559..572O}
{Oguri}, M., {Taruya}, A., \& {Suto}, Y. 2001, \apj, 559, 572

\bibitem[{{Oguri} {et~al.}(2002){Oguri}, {Taruya}, {Suto}, \&
  {Turner}}]{2002ApJ...568..488O}
{Oguri}, M., {Taruya}, A., {Suto}, Y., \& {Turner}, E.~L. 2002, \apj, 568, 488

\bibitem[{{Peiris} {et~al.}(2003){Peiris}, {Komatsu}, {Verde}, {Spergel},
  {Bennett}, {Halpern}, {Hinshaw}, {Jarosik}, {Kogut}, {Limon}, {Meyer},
  {Page}, {Tucker}, {Wollack}, \& {Wright}}]{2003ApJS..148..213P}
{Peiris}, H.~V., {Komatsu}, E., {Verde}, L., {Spergel}, D.~N., {Bennett},
  C.~L., {Halpern}, M., {Hinshaw}, G., {Jarosik}, N., {Kogut}, A., {Limon}, M.,
  {Meyer}, S.~S., {Page}, L., {Tucker}, G.~S., {Wollack}, E., \& {Wright},
  E.~L. 2003, \apjs, 148, 213

\bibitem[{{Pen} {et~al.}(2003){Pen}, {Zhang}, {van Waerbeke}, {Mellier},
  {Zhang}, \& {Dubinski}}]{2003ApJ...592..664P}
{Pen}, U., {Zhang}, T., {van Waerbeke}, L., {Mellier}, Y., {Zhang}, P., \&
  {Dubinski}, J. 2003, \apj, 592, 664

\bibitem[{{Porciani} \& {Madau}(2000)}]{2000ApJ...532..679P}
{Porciani}, C. \& {Madau}, P. 2000, \apj, 532, 679

\bibitem[{{Press} \& {Schechter}(1974)}]{1974ApJ...187..425P}
{Press}, W.~H. \& {Schechter}, P. 1974, \apj, 187, 425

\bibitem[{{Ricotti}(2003)}]{2003MNRAS.344.1237R}
{Ricotti}, M. 2003, \mnras, 344, 1237

\bibitem[{{Schneider} {et~al.}(1992){Schneider}, {Ehlers}, \&
  {Falco}}]{1992grle.book.....S}
{Schneider}, P., {Ehlers}, J., \& {Falco}, E.~E. 1992, {Gravitational Lenses}
  (Gravitational Lenses, XIV, 560 pp.~112 figs..~Springer-Verlag Berlin
  Heidelberg New York.~ Also Astronomy and Astrophysics Library)

\bibitem[{{Spergel} {et~al.}(2003){Spergel}, {Verde}, {Peiris}, {Komatsu},
  {Nolta}, {Bennett}, {Halpern}, {Hinshaw}, {Jarosik}, {Kogut}, {Limon},
  {Meyer}, {Page}, {Tucker}, {Weiland}, {Wollack}, \&
  {Wright}}]{2003ApJS..148..175S}
{Spergel}, D.~N., {Verde}, L., {Peiris}, H.~V., {Komatsu}, E., {Nolta}, M.~R.,
  {Bennett}, C.~L., {Halpern}, M., {Hinshaw}, G., {Jarosik}, N., {Kogut}, A.,
  {Limon}, M., {Meyer}, S.~S., {Page}, L., {Tucker}, G.~S., {Weiland}, J.~L.,
  {Wollack}, E., \& {Wright}, E.~L. 2003, \apjs, 148, 175

\bibitem[{{Yoshida} {et~al.}(2003){Yoshida}, {Sokasian}, {Hernquist}, \&
  {Springel}}]{2003ApJ...598...73Y}
{Yoshida}, N., {Sokasian}, A., {Hernquist}, L., \& {Springel}, V. 2003, \apj,
  598, 73

\bibitem[{{Zentner} \& {Bullock}(2002)}]{2002PhRvD..66d3003Z}
{Zentner}, A.~R. \& {Bullock}, J.~S. 2002, \prd, 66, 43003

\bibitem[{{Zhang}(2004)}]{2004ApJ...602L...5Z}
{Zhang}, T. 2004, \apjl, 602, L5

\bibitem[{{Zhang} {et~al.}(2003){Zhang}, {Pen}, {Zhang}, \&
  {Dubinski}}]{2003ApJ...598..818Z}
{Zhang}, T., {Pen}, U., {Zhang}, P., \& {Dubinski}, J. 2003, \apj, 598, 818

\bibitem[{{Zhao}(1996)}]{1996MNRAS.278..488Z}
{Zhao}, H. 1996, \mnras, 278, 488

\end{thebibliography}
\bibliographystyle{apj}

\appendix

\end{document}